\documentclass[12pt]{article}

\usepackage{amsmath,amssymb}
\usepackage{graphicx}
\usepackage{epstopdf}
\usepackage{epsfig}
\usepackage{array,multirow}

\begin{document}


\title{
SUSY Splits, But Then Returns
}

\author{{ 
{Raman Sundrum\thanks{e-mail: sundrum@pha.jhu.edu}}} \\[0.7cm]
{\it Department of Physics and Astronomy, Johns Hopkins 
University}\\
{\small \it 3400 North Charles St., Baltimore, MD 21218 USA}}

\date{}
\maketitle

\begin{abstract}
We study the phenomenon of accidental or ``emergent'' supersymmetry 
within gauge theory and connect it to the scenarios of Split 
Supersymmetry and Higgs compositeness.  Combining these elements
 leads to a significant refinement and 
extension of the proposal of Partial Supersymmetry, 
 in which supersymmetry is broken at very high energies 
but with a remnant surviving to the weak scale. 
The Hierarchy Problem is then solved 
by a 
non-trivial partnership between supersymmetry and compositeness, 
giving a promising approach for reconciling 
Higgs naturalness with the wealth of precision experimental data. 
We discuss aspects of this scenario from the AdS/CFT dual viewpoint of  
 higher-dimensional warped compactification.
It is argued that 
string theory constructions with high scale supersymmetry breaking which
realize warped/composite solutions to the Hierarchy Problem may 
well be accompanied by some or all of the features described. 
The central  phenomenological considerations and 
expectations are  discussed, with more detailed modelling within 
warped effective field theory reserved for 
future work. 
\end{abstract}

\newpage


\section{Introduction}



In the last decade, models of warped compactification  have
 led to significant progress in understanding 
how non-supersymmetric dynamics can  generate  
the electroweak and flavor hierarchies observed in particle physics.
See Refs. \cite{review} \cite{kkpheno} for reviews and extensive references.
These hierarchies ultimately derive from the 
redshift or ``warp factor'' that varies exponentially across 
extra-dimensional space. While the original Randall-Sundrum I (RS1) \cite{rs1} 
braneworld 
model used such redshifts to solve the electroweak Hierarchy Problem, 
realistic modern variants 
with the Standard Model (SM) 
propagating in 
five-dimensional (5D) ``bulk'' spacetime also
describe flavor hierarchies in terms of 
warped extra-dimensional wavefunction overlaps. 
The most spectacular (but 
challenging)  experimental prediction of such models is the production of
Kaluza-Klein (KK) excitations of gravitons and SM particles at several TeV 
(as reviewed in Ref. \cite{kkpheno}).

Via the  AdS/CFT correspondence \cite{adscft} these warped 
models can be seen as effective descriptions of
 purely 4D
 theories \cite{rscft} 
\cite{nima} \cite{zaff}, 
which however involve strong-coupling physics. 
The extra-dimensional  
warping  that solves the hierarchy problem is dual to 
TeV-scale Higgs compositeness. Weak-coupling in the warped description 
reflects a $1/N_{color}$-type expansion of the 4D strong dynamics.
From the 4D perspective one must posit 
(i)  a weakly-coupled gauge theory sector of elementary particles,
(ii) a strongly-coupled matter sector also charged  under (i), 
with a mass gap $\Lambda_{comp}$, below which lie a finite number 
 of light composites and
above which the dynamics is approximately conformally invariant, and 
(iii) a 
finite list of local
 operators of the strong conformal field theory (CFT)
 needed to couple the two sectors. 
If one further assumes that
 all other minimal color-singlet
CFT operators\footnote{For instance, these are
 single-trace operators in theories with only
color-adjoint ``gluons''.} have large scaling dimensions, greater than some 
$\Delta \gg 1$, then 
$1/\Delta$ emerges as a new expansion parameter for the system.  
The combined $1/\Delta$- and $1/N$-expansion is 
 precisely  a weakly-coupled 5D 
warped effective field theory on an RS1 background. More complex situations 
can also arise with even higher dimensional warped descriptions 
(famously the case in
  ${\cal N} =4$ supersymmetric Yang-Mills \cite{adscft}), or without 
any higher-dimensional 
effective field theory regime. In this sense, 5D warped models 
are conjectured minimal realizations of (i -- iii), but they display
 many qualitative features that appear to be robustly general. 

Warped models represent a powerful and
comprehensive approach to the hierarchy problem and other challenges
 of particle physics, providing an attractive alternative to
 the paradigm of Weak Scale Supersymmetry (SUSY). But there are 
two important regards in which they are not completely satisfactory.   
The first is theoretical, in that higher-dimensional effective field theory 
is non-renormalizable and warping pushes this feature to
relatively low energies.
 Even though many leading effects, even quantum effects, 
are calculable in such a framework, they ultimately rely on the {\it 
existence} of a UV completion.  At present, superstring theory provides 
the main hope for finding such UV completions of 
minimal or non-minimal warped models. But since we are trying to explore 
a non-SUSY solution to the hierarchy problem, superstring constructions 
must contend with breaking SUSY at very high scales 
while still retaining control of the rather complex dynamics.
As a result this subject is still 
in its infancy, with constructions aiming at proofs of principle rather than 
full realism \cite{strassler} \cite{shamit}.

The second issue is 
phenomenological. Any resolution of the hierarchy problem
by new TeV-scale physics must explain why that new physics has not been seen
in direct searches or through its virtual effects in precision tests.  
Warped models are in a sense highly successful in this regard, but  still 
imperfect, in that evading precision tests require pushing the new physics
(KK excitations dual to $\Lambda_{comp}$-scale composites) 
to several TeV, leading to percent-level or worse
residual fine-tuning to account for the smaller electroweak scale. 
Analogous tensions are seen in other
approaches to the hierarchy problem 
and are collectively known as the Little Hierarchy Problem. 

The purpose of the present paper is to study the joint role SUSY might play 
in each of these two issues, without however stealing the entire spotlight. 
The plot is as follows. 
It may well be true that SUSY is a fundamental symmetry of spacetime that 
appears in string theory UV completions of warped models of particle physics,
but with very large SUSY-breaking scale. The subtleties of SUSY breaking and
transmission in warped settings is therefore important to understand. 
One such subtlety is that despite high-scale SUSY-breaking,
 vestiges of SUSY can be ``accidentally'' redshifted (warped) 
down from high scales to low scales $\sim \Lambda_{comp}$.  
This can be 
seen in both string and effective field theory examples   \cite{markus1} 
\cite{giddings} \cite{partial} \cite{markus2} \cite{strassler}  \cite{shamit}. 
From the dual 4D 
viewpoint this corresponds to SUSY being an accidental or ``emergent'' 
symmetry of the strongly-coupled sector. This vestigial or ``partial SUSY''  
is insufficient by itself to solve the ``Big''
 Hierarchy Problem, which is instead
solved by warping/compositeness as in non-SUSY models, but it has the power to 
neatly address the Little Hierarchy Problem by stabilizing a modestly larger 
$\Lambda_{comp}/$weak-scale hierarchy than naturally allowed in non-SUSY 
models. 
Such a mechanism is not  
merely a theoretical nicety: the physics that resolves the Little 
Hierarchy Problem will dominate particle physics experiments in the years to 
come, in particular the Large Hadron Collider (LHC).

We build on the work of Ref. \cite{partial}, which (in 4D language) 
 proposed the co-existence of emergent SUSY of the
strong sector  composites  with the absence of SUSY in the  
weakly-coupled sector. They argued that this Partial SUSY could be
 protected 
at strong coupling in a manner
 dual to the extra-dimensional separation of the two 
sectors in a warped compactification. With the Higgs bosons and top quarks 
among the SUSY composites, partial SUSY would in turn protect the 
``little hierarchy'' 
between $\Lambda_{comp}$ and the weak scale. 
But while top quark loops give the largest Higgs
 radiative corrections  that destabilize the weak scale, 
gauge loops are not far behind. 
We show that there is a more powerful extension of partial SUSY that follows 
by replacing the completely non-SUSY weakly-coupled sector by a 
weakly-coupled sector of Split-SUSY \cite{split1} \cite{split2} \cite{split3}
 type, in particular 
containing light gauginos.  
This extension  then suppresses all the dominant 
Higgs radiative corrections in the little hierarchy.\footnote{
It is important to distinguish our proposal from that of
 section 3.5 of Ref. \cite{split1}, which also combines Split SUSY with 
a strong CFT sector at the TeV scale. The central difference is that 
the CFT of Ref. \cite{split1} does not have any (accidental) SUSY, 
and there is no extra suppression of the dominant Higgs
 corrections beyond that offered by Higgs compositeness. }
 A  new analysis 
is given of 
the requisite SUSY cancellations in gauge loops, taking into account that the
gauginos and gauge   particles are elementary, unlike
 the Higgs-top-stop system which consists entirely of low-scale composites.

We also carefully discuss instabilities in the proposal of partial SUSY 
which arise
 from  AdS-tachyons which are scalar superpartners of 
gauge fields in the warped bulk. While such tachyons do not 
signify instabilities of infinite 
AdS spacetime, they do
 represent potential 
instabilities of warped compactifications after high-scale 
SUSY breaking \cite{strassler}. 
Indeed, attempts at string 
UV-completion must inevitably deal with some version of this issue, 
and it explains the challenge in finding such constructions.
Here we study the issues in their most minimal setting.
We will show how such instabilities can be avoided by 
extending the SM gauge symmetry, resulting in new multi-TeV 
gauge bosons with order one couplings to light quarks and leptons,
just possibly in range of the LHC.

In this way, we will describe a partnership in establishing the electroweak 
hierarchy, between SUSY and warping/compositeness. SUSY may be in charge at the
very highest scales, with warping/compositeness then generating and protecting 
a significant hierarchy, and then partial SUSY returning at the TeV scale 
to generate a little hierarchy.  Phenomenologically, the lowest rung of this 
ladder has similar collider implications to 
 More Minimal SUSY \cite{moremin},  followed by multi-TeV
 gauge bosons and KK-excitations/massive composites.  
Most of our discussion in this paper
 will be given in  4D language, in which the basic ``grammar'' 
of the story can be worked out using the power of the 
renormalization group (RG) to connect disparate scales. This will lay the 
foundation for Ref. \cite{future}, 
in which minimal warped models will be built 
and more quantitatively analysed.

The paper is organised as follows. In Section 2, we discuss 
the phenomenon of accidental SUSY within gauge theory 
and its connection to strongly coupled matter.
In Section 3, we review a simple structure of
 UV SUSY breaking compatible with 
Split SUSY. In Section 4, we review how a mass gap, $\Lambda_{comp}$, can 
arise in the IR of a strong CFT and then study how effective the associated 
physics is at  mediating SUSY breaking effects to the IR.
We  study how light
composites, in particular scalars,  of the accidentally 
supersymmetric strong dynamics feel UV SUSY breaking.
 The discussion in this section is restricted to the 
weakly-coupled sector being (accidentally) pure SUSY Yang-Mills theory.  
Section 5 discusses an accidentally SUSY strong sector 
coupled to a Split SUSY weakly-coupled sector. This is our 
final target, yielding our improved version of partial SUSY. 
We focus on how naturalness of composite scalars works 
in this setting. In Section 6, we give a lightning
 CFT-dual review of realistic 
non-SUSY warped models. In Section 7, we then incorporate partial SUSY and 
focus on several of the key issues for  realistic model-building.
In Section 8, we briefly discuss the experimental implications of 
our scenario. 

This work grew out of an initial collaboration with
Thomas Kramer,  
investigating the possibility of realistic partial SUSY \cite{tom}.
However, the proposal  to 
fuse partial SUSY with Split SUSY is new to the present paper, 
as are a number of lesser issues. The ideas of the present paper are
quite distinct from the ``partial split SUSY'' scenario of Refs. \cite{partialsplit}.

\section{Accidental Supersymmetry within Gauge Theory}

IR fixed points readily appear in 
 the RG flows of 
  quantum effective field theories.  Consider such a fixed point 
with a robust basin of attraction. That is, after imposing 
some symmetries in the UV, for a non-fine-tuned set of 
UV couplings the theory flows towards the fixed point, at least
 over a large range of energies.
The CFT describing the fixed point 
may possess greater symmetry than is present in the UV, in which case 
the enhanced symmetry is  
 referred to as ``accidental'' or ``emergent''.
The fixed point is at best a limit of the RG flow, so accidental symmetries 
are never exact but become better approximations the further the 
 IR running towards the fixed point. Here, we study the case of 
accidental supersymmetry \cite{kaplan} in the context of gauge theories with 
perturbatively weak gauge coupling. Refs. \cite{markus2} \cite{strassler}
 discussed the case where there is no weakly-coupled gauge theory outside the
strong dynamics.\footnote{In Ref. \cite{markus2}, weakly-coupled gauge fields 
emerge in the IR as composites of the strong dynamics.}

\subsection{Accidental SUSY Yang-Mills theory}

Pure SUSY Yang-Mills (SYM) offers the only completely weak-coupling example of 
accidental SUSY in gauge theory \cite{kaplan}.\footnote{
Ref. \cite{kaplan} {\it derives} the particle 
content of SYM  in the IR using strong dynamics in the UV.
 In this subsection we 
simply assume the right particle content and review how SUSY couplings emerge 
in the IR. In this way we avoid the strong coupling regime of Ref. 
\cite{kaplan}.}
 Suppose that in the far UV, SUSY is 
either non-existent or badly broken, so that at lower energies we have a 
non-SUSY gauge theory consisting of a gauge field and  a Weyl fermion, 
$\lambda^a$, in 
the adjoint representation of the gauge group. Let us assume that the UV 
dynamics respects the chiral symmetry $\lambda \rightarrow e^{i \theta} 
\lambda$ (more precisely, a non-anomalous discrete subgroup of this 
symmetry) so that a fermion mass term is  forbidden. Then the effective gauge 
theory takes the form
\begin{equation} 
{\cal L}_{eff}(M) = - \frac{1}{4} F_{\mu \nu}^2 + \bar{\lambda} i D. \sigma 
\lambda + \frac{\rm higher~dimension~  operators}{M^k},
\end{equation}
where $M$ is the scale of massive physics we have integrated out.
Since the higher-dimension operators are not constrained to respect SUSY, 
at scales just below $M$ the theory is far from supersymmetric. But in the 
far IR the theory is well approximated by just the {\it renormalizable} 
gauge and chirally invariant interactions, 
\begin{equation} 
{\cal L}_{eff}(\mu \ll M) 
= - \frac{1}{4} F_{\mu \nu}^2 + \bar{\lambda} i D. \sigma 
\lambda,
\end{equation}
which is accidentally supersymmetric. In this way, SUSY has emerged or 
re-emerged in the IR. 
Identifying the IR theory as approximate 
SYM, $\lambda$ is identified with the 
 gaugino and its chiral symmetry with $R$-symmetry. 

Note that for any particular starting UV couplings, there
is a limit to how far we can run towards supersymmetry in the IR, given by 
the mass gap or confinement scale, ``$\Lambda_{QCD}$'', but 
this can naturally be $\ll M$. Physically, the spectrum of
 ``hadrons'' will 
not be exactly supersymmetric, but will have small 
SUSY-breaking splittings of order $\Lambda_{QCD}/M$, due to the 
residual sensitivity to non-renormalizable interactions.

\subsection{Charged matter via strong coupling}

Let us try to 
generalize to accidental SUSY gauge theories with  charged matter. 
It is immediately 
obvious that examples do not exist at weak coupling. Any such 
example beyond SYM would result in chiral 
supermultiplets (in ${\cal N} =1$ SUSY language) 
which contain scalars. Whatever the   
quantum numbers of such scalars $\phi$, with the exception of SUSY which we 
take to be broken or non-existent in the UV, we cannot naturally 
forbid a completely 
symmetric mass term,
\begin{equation}
{\cal L}_{eff}(M) \ni - M^2 |\phi|^2,
\end{equation}
 and without fine-tuning $\phi$ will not survive 
in the IR.

But there is an important 
loop-hole to this argument at strong coupling \cite{lutrat}, 
since large anomalous dimensions might give 
the completely symmetric $|\phi|^2$ operator a true scaling 
dimension $>4$, thereby 
making it an {\it irrelevant} SUSY-breaking perturbation, that flows towards 
zero in the IR. Note that we are assuming the gauge dynamics discussed
above is {\it weakly-coupled}, and that the strong coupling is due to some 
other interactions of the charged matter (possibly gauge 
interactions associated with a different gauge group). 
At first sight, the possibility that strong interactions can make 
a highly-relevant SUSY-breaking mass term into an irrelevant effect
 might seem a far-fetched hope, in any case 
untestable in the face of the difficulties of strong coupling 
calculations. But remarkably, strongly-coupled large-$N_{color}$ 
${\cal N}=4$ supersymmetric
Yang-Mills theory demonstrates precisely this mechanism \cite{strassler}! 
The gauge scalars of 
this theory, have a completely gauge- and R-symmetric bilinear operator 
$|\phi|^2$, which has a scaling dimension of order $N_{color} 
\gg 4$ (at strong 
coupling), thereby 
making it an irrelevant SUSY-breaking deformation of the theory. 
This is established
 by examining  the AdS/CFT dual string description.

Having seen this 
strong-coupling ``miracle'' occur under theoretical control, 
we should consider that it might well be a robust phenomenon 
among strongly-coupled theories, 
and might play an important role in the real world. 
Let us therefore further 
explore the possibility of accidental SUSY in weakly-coupled 
gauge theory with strongly-coupled matter. 

\subsection{Minimal set of relevant and marginal couplings}

We consider the possibility of 
a strongly-coupled matter sector that flows towards 
a supersymmetric fixed point, described by a superconformal field theory 
(SCFT), weakly coupled to a set of gauge fields and ``accidental 
gauginos'' (in the sense described in subsection 2.1).
 Let us be maximally optimistic and assume that the fixed point is 
as attractive as possible in the IR. That is, we assume the SCFT has the 
fewest possible 
relevant scalar operators that might appear in the interactions of the 
UV  theory and push it away from the SUSY fixed point in the IR. 
If there are no highly relevant scalar operators we will have accidental 
SUSY. We want to check whether this is consistent with the general
structure of a SCFT of gauge-charged matter. 

Every CFT comes with an energy-momentum tensor
$T_{\mu \nu}$ with scaling dimension $4$. Because it is conserved and 
traceless, it contains no  scalar component. In a SCFT, 
there must also be a vector current related to $T_{\mu \nu}$ by SUSY,
but again it is conserved and contains no scalar component, and is just the 
symmetry current of the  superconformal $R$-symmetry. Thus the absolutely 
minimal structure of a SCFT does not imply on general grounds 
the existence of scalar operators 
with ${\cal O}(1)$ scaling dimensions \cite{markus2}.

In 
addition, we want the 
SCFT matter to be weakly gauged by some external gauge fields 
and ``gauginos''.
 This means that prior to being 
gauged, the SCFT possesses a non-$R$ symmetry, with conserved currents
$J_{\mu}^a$, again without any scalar component. But now SUSY necessitates
 scalar operators $D^a$ in the same supermultiplet as $J_{\mu}^a$. 
These are nothing but ``$D-terms$'', which are the familiar 
 simple ``squark'' bilinears 
in weakly coupled matter sectors, but at strong coupling
 such identifications become less useful and we work directly with the 
$D^a$ operators and their symmetry and scaling properties. 
Conserved currents  $J_{\mu}$ have protected scaling dimension $3$, 
and SUSY then implies that the $D^a$ have protected scaling dimension $2$.
We must therefore worry about relevant deformations of the UV theory of the 
form
\begin{equation}
{\cal L}(M) \ni {\cal L}_{SCFT} + M^2 D,
\end{equation}
which would prevent it from flowing towards the supersymmetric fixed point, 
without fine-tuning.
Clearly however, for non-abelian gauge symmetry, this is forbidden by 
gauge invariance itself. So we must only consider abelian factors of the 
gauge group.  Even here, exact discrete symmetries such as charge-conjugation
invariance can forbid such couplings linear in $D$'s. We assume for 
the rest of this section that such protective symmetries are in place. 
In connecting 
to the real world this provides a tight model-building requirement as we 
discuss in Section 7.

Thus the contrast between strongly-coupled and weakly-coupled SCFT 
limits of charged matter is this. Both contain 
dimension-$2$ scalar operators corresponding to potentially relevant 
SUSY-breaking deformations of the SCFT, but 
it is only weakly coupled SCFTs that 
necessarily possess dimension-$2$ scalar operators invariant 
under all (non-SUSY) symmetries, that cannot naturally 
be forbidden from appearing 
in the UV. 

Let us now assemble the minimal set of couplings of the SCFT matter and 
the external gauge fields, $A_{\mu}^a$, and gauginos, $\lambda^a$. 
We will do this without 
regard to SUSY, just imposing gauge invariance (and any protective 
non-SUSY symmetries for abelian gauge groups and massless gauginos), 
and keeping all couplings which are 
relevant or marginal in the IR. We will then study the extent to which these
couplings flow towards SUSY relationships in the IR. The most obvious 
of these is the gauge coupling, $g$, of  gauge fields to themselves (if 
non-abelian) and also to SCFT matter. 
The SCFT supermultiplet containing the conserved current $J_{\mu}^a$, 
and the scalar operator $D^a$,  also contains a fermionic operator 
$\Psi_J^a$ with protected scaling dimension $5/2$. These can couple 
gauge-invariantly and marginally to the $\lambda^a$.

So far we have only 
considered couplings linear in the conserved currents of the SCFT or 
in operators in the same supermultiplets. But we must also consider couplings
involving Lorentz-invariant {\it products} of such operators. 
 In general however, the relevance of 
such operator products is difficult to establish since their scaling 
dimensions need not be algebraically related to those of 
their factors. We will therefore specialize to the case of strongly-coupled 
SCFTs with a large-$N$ expansion, in which case the scaling 
dimension of a product is given  simply by the sum of the dimensions of its 
factors, up to $1/N$ corrections. We will denote the associated parameter 
``$N$'' from now on as ``$N_{CFT}$'' in order to be clear that it 
 does not refer to the size of the weakly coupled gauge 
sector external to the CFT, which is taken to be parametrically smaller.
In such an expansion, we see that the only 
scalar product operator of the SCFT that need concern us is the approximately 
marginal 
(and completely symmetric) product $D^a D^a$. The effective Lagrangian 
describing the RG flow is therefore given simply by
\begin{equation}
{\cal L}(\mu) = -\frac{1}{4} F_{\mu \nu}^{a~2} + \bar{\lambda} D.\sigma \lambda
+ {\cal L}_{SCFT} + g A_{\mu}^a J^{\mu}_a + \tilde{g} \lambda_a \Psi_J^a 
- \frac{1}{2} g_D^2 D^a D^a.
\end{equation}
Here the term ``$g A_{\mu}^a J^{\mu}_a$'' is schematic. It means that 
the SCFT is gauged by the external gauge group, with coupling $g$. 
In general this is 
a non-linear coupling to $A_{\mu}$ determined by gauge invariance, but at 
linearized order it takes the above form. Note that we are insisting on 
the UV   
chiral symmetry of the ``gaugino'' $\lambda$, protecting against 
a mass term $\lambda \lambda$. 
In this way, 
our RG flow is parametrized by just three (approximately) marginal 
 couplings, $g, \tilde{g}, g_D^2$.

\subsection{RG in the large-$N_{CFT}$ expansion} 

Let us start by  working out the RG  equations for $g, \tilde{g}, g_D^2$, 
to one-loop order in these weak couplings, and to leading order in the 
large-$N_{CFT}$ expansion of the strong interactions of the SCFT sector.

 The gauge coupling running is dominated 
by the large-$N_{CFT}$ matter in vacuum polarization, 
\begin{equation}
\frac{d g^2}{d \ln \mu} = \frac{{\cal O}(N_{CFT})}{16 \pi^2} g^4.
\end{equation}
The fermionic operator $\Psi_J$ in the $\tilde{g}$ coupling
 is not renormalized purely by the 
SCFT dynamics since it is related to the conserved current $J_{\mu}$ by SUSY. 
It can be corrected by further $g, \tilde{g}, g_D^2$ couplings, but without 
$N_{CFT}$-enhancement. However, $\tilde{g}$ can be renormalized by wavefunction
renormalization of the gaugino $\lambda$ by the SCFT, which  is 
$N_{CFT}$-enhanced. 
Note this involves two more $\tilde{g}$ couplings, so that 
\begin{equation}
\frac{d \tilde{g}^2}{d \ln \mu} = \frac{{\cal O}(N_{CFT})}{16 \pi^2} 
\tilde{g}^4.
\end{equation}
Similarly, in $g_D^2$ renormalization, the factors of $D^a$ are not 
renormalized purely by the SCFT dynamics at the fixed point, 
$D^a$ being related by SUSY to the conserved current, 
and corrections to this from external interactions are not $N_{CFT}$-enhanced.
 Instead, the large-$N_{CFT}$ enhanced renormalization comes at order
$(g_D^2)^2$, from contractions of the form $g_D^4 
D^a \langle D^a D^b \rangle D^b$. Therefore, 
\begin{equation}
\frac{d g^2_D}{d \ln \mu} = \frac{{\cal O}(N_{CFT})}{16 \pi^2} g_D^4.
\end{equation}

Note that if we had exact SUSY, it would require
\begin{eqnarray}
{\cal L}(\mu) &=& {\cal L}_{SCFT} + \int d^2 \theta ~{\cal W}_{\alpha}^a {\cal W}^{\alpha}_a  
+ g \int d^4 \theta  ~V_{gauge} J_{CFT}  \nonumber \\
&=& {\cal L}_{SCFT} - 
\frac{1}{4} F_{\mu \nu}^{a~2} + \bar{\lambda} D.\sigma \lambda
+  g A_{\mu}^a J^{\mu}_a + g \lambda_a \Psi_J^a 
- \frac{1}{2} g^2 D^a D_a, 
\end{eqnarray}
where in the second line we have integrated out the scalar auxiliary 
superpartner of the gauge field.\footnote{Off-shell, this auxiliary 
field  is distinct 
from the composite operator of the matter sector,
$D^a$, but becomes equal to it after solving its equation of motion.}
That is, exact SUSY would require
\begin{equation}
 \tilde{g}^2 = g_D^2 = g^2. 
\end{equation}
We do not 
insist on SUSY in the UV, but  
we want to see if the theory flows towards this relation
in the IR. But the RG flow must preserve the SUSY relations if they 
were satisfied in the UV. This tells us that the 
${\cal O}(N_{CFT})$ coefficients of each of the 
RG equations for $g^2, \tilde{g}^2, g_D^2$ is the {\it same}, so that
\begin{equation}
\label{ldrg}
\frac{d 1/g^2}{d \ln \mu} = \frac{d 1/\tilde{g}^2}{d \ln \mu} = 
\frac{d 1/g^2_D}{d \ln \mu} \equiv 
 - b_{CFT} \sim \frac{{\cal O}(N_{CFT})}{16 \pi^2}.
\end{equation}

Of course, these same RG equations govern the flow of couplings even 
when they are non-supersymmetrically related. Let us assume that 
at some UV scale, $m_0$, we start with three unrelated non-SUSY couplings, 
$g_0, \tilde{g}_0, g_{D 0}^2$. In the IR, these flow to 
\begin{equation}
g^2 = \frac{g_0^2}{1 + b_{CFT} g_0^2 \ln(m_0/\mu)}; ~ ~ 
 \tilde{g}^2 = \frac{\tilde{g}_0^2}{1 + b_{CFT} \tilde{g}_0^2 \ln(m_0/\mu)};
~ ~ g_D^2 = \frac{g_{D 0}^2}{1 + b_{CFT} g_{D 0}^2 \ln(m_0/\mu)}.
\end{equation}
Working to first order in the splittings of the $g$'s,
\begin{eqnarray}
\frac{\Delta \tilde{g}^2}{g^2} &\equiv& \frac{\tilde{g}^2 - g^2}{g^2} 
\approx \frac{g^2}{g_0^2} \frac{\Delta \tilde{g}^2_0}{g^2_0} \nonumber \\
\frac{\Delta g_D^2}{g^2} &\equiv& \frac{g_D^2 - g^2}{g^2} 
\approx \frac{g^2}{g_0^2} \frac{\Delta g^2_{D 0}}{g^2_0}.
\end{eqnarray}
Since the gauge coupling running is dominated by SCFT matter, it is IR free, 
$b_{CFT} > 0$, and $g$ falls with RG scale $\mu$. Therefore, we see that 
there is a focussing effect for the splittings in the IR. 
Even one hundred percent
 differences among the $g_0^2, \tilde{g}_0^2, g_{D 0}^2$, can still evolve 
to small fractional differences in the IR.  
This is the central realization of  accidental SUSY within gauge theory.

\subsection{Important RG corrections}

The above analysis is formally leading in the large-$N_{CFT}$ limit, but 
it misses a qualitatively important correction, namely the effects 
which are leading for small $g$. If one flows sufficiently far into the IR
such effects will always dominate given the IR-free nature of the gauge 
coupling, driven by the large amount of SCFT matter.

This is not an issue for the  $\beta$-functions 
of $g^2, \tilde{g}^2$, which necessarily start at ${\cal O}(g^4)$ (where we 
are taking any of the $g^2, \tilde{g}^2, g_{D}^2$ to be roughly 
comparable) because 
renormalization occurs either via self-energy corrections to gauge fields 
or gauginos, in which the claim is obvious, or via vertex corrections to 
the $\lambda \psi_J$ vertex. In this latter case, an ${\cal O}(g^2)$ 
correction to the $\psi_J$ operator is required for non-trivial 
renormalization because the supersymmetric strong dynamics at the 
fixed point alone
does not renormalize this ``superpartner'' of the conserved current $J_{\mu}$.

However, the $g_D^2$ $\beta$-function does 
get an ${\cal O}(g^2)$ correction. First of all there is a possible 
multiplicative renormalization of $g_D^2$, due to 
strong dynamics dressing of the $D^a D^a$ operator. But it is straightforward
to see that this is a $1/N_{CFT}$-suppressed effect since it involves 
connecting two separate CFT-color singlet factors. Gauge field
and gaugino exchanges alone cannot additively renormalize $g_D^2$, because 
their spin forbids their producing the correct scalar-scalar coupling of 
strong operators, and because such exchanges are not 1PI. Both of these 
deficiencies however are corrected by further strong dynamics dressing, 
but again only in a $1/N_{CFT}$-suppressed manner. Consequently, 
the RG equation for $g_D^2$ is modified to the form
\begin{equation}
\frac{d g_D^2}{d \ln \mu} = b_{CFT} g_D^4   + {\cal O}(1/N_{CFT}) g_D^2 
+  {\cal O}(1/N_{CFT}) g^2 + {\cal O}(1/N_{CFT}) \tilde{g}^2.
\end{equation} 
Again, the various ${\cal O}(1/N_{CFT})$ coefficients must be related 
in such a way as to preserve the SUSY relationship $g^2=\tilde{g}^2=g_D^2$ 
if it holds in the UV.
Working to first order in $\Delta g_D^2, \Delta \tilde{g}^2$, this implies 
\begin{equation}
\frac{d \Delta g_D^2}{d \ln \mu} = 2 b_{CFT} g^2 \Delta g_D^2   
- \gamma_D \Delta g_D^2 - \tilde{\gamma} \Delta \tilde{g}^2, 
\end{equation}
where we have new constants,
\begin{equation}
\gamma_D, \tilde{\gamma} \sim {\cal O}(1/N_{CFT}).
\end{equation}

We see that the $\gamma$ terms can become more important in the IR 
because of IR freedom of the gauge coupling and the scaling of these terms as 
${\cal O}(g^2)$ rather than ${\cal O}(g^4)$. In more detail, we can simplify 
by noting 
 that the flow of
 $\Delta \tilde{g}^2$ is unaffected and therefore our earlier 
derivation is still valid, namely that it actually 
scales as ${\cal O}(g^4)$ in the IR.
Thus
we can consistently neglect its feedback into $\Delta g_D^2$ via $\tilde{\gamma}$ in 
the IR flow. The dominant danger to our large-$N_{CFT}$ analysis comes 
from $\gamma_D$. Keeping just this new effect, 
the full set of RG equations is given by,
\begin{eqnarray}
\frac{d g^2}{d \ln \mu} &=& b_{CFT} g^4  \nonumber \\
\frac{d \Delta \tilde{g}^2}{d \ln \mu} &=& 2 b_{CFT} g^2 \Delta 
\tilde{g}^2  \nonumber \\
\frac{d \Delta g_D^2}{d \ln \mu} &=& (2  b_{CFT} g^2    
- \gamma_D) \Delta g_D^2.
\end{eqnarray}

These corrected RG equations are still straightforward to solve,
\begin{eqnarray}
\label{deltasol}
g^2 &=& \frac{g_0^2}{1 + b_{CFT} g_0^2 \ln(m_0/\mu)} \nonumber \\
\frac{\Delta \tilde{g}^2}{g^2}  &=&
 \frac{g^2}{g_0^2} \frac{\Delta \tilde{g}^2_0}{g^2_0} \nonumber \\
\frac{\Delta g_D^2}{g^2}  &=&
 \frac{g^2}{g_0^2} \frac{\Delta g_{D 0}^2}{g^2_0} 
(\frac{m_0}{\mu})^{\gamma_D}.
\end{eqnarray}
If $\gamma_D > 0$ 
we cannot trust these solutions arbitrarily into the IR. At some point, 
even though $\gamma_D \sim {\cal O}(1/N_{CFT})$, power law growth of 
$\Delta g_D^2$ in $\mu$ will take over, and it will be necessary 
to treat $\Delta g_D^2$ beyond first order. In the remainder of the paper 
we will take care that we are in the linear regime. 

We have just discussed subleading effects in 
large-$N_{CFT}$ that are of order ${\cal O}(g^2)$. 
There remains one last class of effects in the RG equations that appears at 
one-loop order in the $g^2, \tilde{g}^2, g_D^2$ deviations from the 
exact fixed point SCFT dynamics, namely all the subleading effects in 
large-$N_{CFT}$ that are of order ${\cal O}(g^4)$. 
The simplest of these is just the running contribution to the gauge coupling
due to gauge fields and gauginos themselves when the external gauge group is 
 non-abelian, 
\begin{equation}
\frac{d 1/g^2}{d \ln \mu} = - b_{CFT} + b_{SYM}.
\end{equation}
For example, for an $SU(n)$ external gauge group ($n$ formally smaller 
than $N_{CFT}$), $b_{SYM} = 3n/(8 \pi^2)$. 
But the new complication at this order is that 
the linearized RG equations for 
$\Delta g_D^2$ and $\Delta \tilde{g}^2$ will in general be
coupled to each other, and consequently harder to solve. However, these 
corrections are truly subdominant to the effects calculated above. 
In this paper, we will neglect them.

The power of  minimal warped compactification models
related to this scenario by AdS/CFT, is that 
the entire set of RG corrections discussed above
 is calculable in terms of just the one strong interaction parameter, 
$b_{CFT}$, and group theory considerations.
We will do a more complete analysis of these corrections in Ref.\cite{future}. 
Below we shall just illustrate the value-added in minimal warped models
 by showing how  $\gamma_D$ is determined in terms of $b_{CFT}$.
In particular we verify that $\gamma_D >0$, so that 
it indeed signifies a (slightly) relevant  effect in the IR.

\subsection{$\gamma_D$ in minimal warped compactification}

Let us translate some of the issues surrounding the SCFT operators
$D^a$ and  $D^a D^a$ into the context of warped compactifications, using 
AdS/CFT. The fact that the CFT is gauged by an external gauge group, 
means in isolation it must contain corresponding conserved currents 
$J_{\mu}^a$. The dual of these currents and the external gauge fields 
is given by 5D gauge fields propagating on a 5D warped RS 
background. 
 SUSY at the fixed point translates into minimal SUSY in the 5D bulk, which 
means that the bulk minimally also contains 5D gauginos and 5D gauge scalars.
It is these gauge scalars which are dual to the $D^a$. The scaling 
dimension $2$ of the $D^a$ matches the AdS-SUSY constraint that the 
5D gauge scalars are AdS tachyons saturating the Breitenlohner-Freedman 
bound \cite{bf}.  This set-up is our 
minimal bulk system. 5D gravity, dual to the CFT energy-momentum tensor, can 
naturally play a subdominant role (for sufficiently large 5D Planck mass). 
We are imagining that SUSY is violently 
broken or absent on the UV boundary, but that the rest of the dynamics is 
supersymmetric. 

The bulk 5D SUSY Yang-Mills dyamics is controlled by a single parameter,
the $5D$ gauge coupling, $g_5^2$, which can be expressed  dimensionlessly, 
$g_5^2/R_{AdS}$, in terms of the AdS radius of curvature. We can identify 
it within the SCFT description by doing a tree-level matching from 5D to the 
IR 4D gauge coupling below the KK scale:
\begin{eqnarray}
\frac{1}{g^2} &=& \frac{1}{g_0^2} + \frac{L}{g_5^2} \nonumber \\
&=& \frac{1}{g_0^2} + \frac{\ln(m_0/\Lambda_{comp}) R_{AdS}}{g_5^2}.
\end{eqnarray}
Here, $1/g_0^2$ is the coefficient of the UV 
boundary-localized kinetic term for 
the gauge fields, 
$L$ is the length of the fifth dimension, $m_0$ is the UV scale
and $\Lambda_{comp} = m_0 e^{-L/R_{AdS}}$ is the resulting warped down 
IR scale.  AdS/CFT identifies this equation with the RG equation for
the external gauge coupling at leading order in large $N_{CFT}$. We thereby 
arrive at the identification,
\begin{equation}
b_{CFT} \equiv \frac{ R_{AdS}}{g_5^2}.
\end{equation}

While AdS tachyons at or above the Breitenlohner-Freedman bound, such as 
our gauge scalars,  do 
not represent an instability of AdS, they do represent a violent instability 
in an RS background if they are allowed tadpole couplings on the 
UV boundary, since these lead to scalar profiles that rapidly blow up towards 
the IR of the warped bulk. This is precisely why we needed to introduce 
protective symmetries in the UV. In the 5D picture these symmetries forbid
one from writing or radiatively generating such a tadpole 
couplings in the UV. In the dual picture they protect against deformations of 
the SCFT linear in the $D^a$.

However, one must also consider quantum loops of the tachyons in the 
5D theory, with effects analagous to  Casimir energy in a compact 
space. Such loops allow tachyon-pairs to propagate into the IR of the 
warped bulk, dual to the effects of deforming the SCFT by $D^a D^a$. 
 These loops cannot be forbidden by 
any symmetry. SUSY can enforce their cancellation but we have taken it to be
 maximally
 broken on the UV boundary, say by   writing
UV-boundary localized mass terms for the 5D gauge scalars. 
We are interested in seeing how efficiently 
this UV SUSY breaking propagates into the IR as a result of the tachyon loops.
 At leading order the tachyons are free particles in the bulk, dual to 
 the leading large-$N_{CFT}$ approximation, in which 
the operator $D^a D^a$ simply has twice the dimension of $D^a$.
That would make it a marginal dimension-$4$ deformation. However, 
minimally, the gauge scalars (for non-abelian gauge group) will 
exchange gauge bosons in the bulk. It is precisely this effect that 
is dual to the $\gamma_D$ correction in the SCFT. 

We can compute it efficiently (following a similar strategy to 
Refs. \cite{gamma}) by studying the 
SCFT correlator
\begin{eqnarray}
\langle D^a(0) D^a(x) D^b(x') D^c(x'') \rangle &\ni_{x \rightarrow 0}& 
\frac{\langle (D^a D^a)(0) D^b(x') D^c(x'') \rangle}{|x|^{\gamma_D}} \nonumber \\
&\approx_{\gamma_D \rightarrow 0}& 
\langle (D^a D^a)(0) D^b(x') D^c(x'') \rangle \times (1 - \gamma_D \ln|x|).
\end{eqnarray}
The first line is part of an OPE, with the exponent of $x$ following by 
dimensional analysis using the scaling dimensions of all the local operators. 
The second line is expanded formally for small 
$\gamma_D$. We know that $g_5^2/R_{AdS} = 
1/b_{CFT} \propto 1/N_{CFT}$, formally the same order as $\gamma_D$. 
Therefore perturbative $AdS_5$ calculations of the above correlator 
in $g_5^2$ are automatically 
perturbative expansions in $\gamma_D$. No finite set of Yang-Mills 
Feynman diagrams in AdS will reproduce the first line, but the leading orders
will give us the second line and we can then ``RG improve'' the result to get 
the first line. In fact all we want to extract is $\gamma_D$, which requires 
us to compare the small $x$ asymptotics of the original correlator at 
leading (zeroth order in $g_5^2$) and next-to-leading (order $g_5^2$). 
This translates into the AdS correlator of four gauge scalars by the standard 
AdS/CFT prescription, with leading order corresponding to free field theory 
approximation for the scalars, and next-to-leading order corresponding to 
single-gauge-boson exchange between the scalars in 5D. The first of these 
is straightforward, while the latter can be extracted from the work of 
Ref. \cite{gamma2}. Assembling the result gives 
\begin{equation}
\label{gamma}
\gamma_D = \frac{n g_5^2}{12 \pi^2 R_{AdS}} \equiv \frac{n}{12 \pi^2 b_{CFT} },
\end{equation}
if the gauge group is $SU(n)$, and zero if it is abelian.

We will use this relation in Section 7 in making some numerical estimates.

\section{R-Symmetric UV SUSY breaking} 

We now want to move to considering SUSY as a fundamental symmetry of the 
far UV, but a badly broken one.
We can capture the high scale 
breaking of fundamental SUSY, or indeed its complete absence, in a simple way.
We assume  that SUSY breaking originates from some hidden sector dynamics
with vacuum energy $V_0$,  
and is communicated 
to  our gauge and matter
sectors by some massive physics of typical mass scale $M$. 
Integrating out this massive physics, 
and with sufficient non-SUSY symmetries in the UV (including the 
$R$-symmetries that protect against gaugino masses), 
 $V_0$ is the only SUSY breaking ``spurion'' felt by the
gauge and matter sectors, suppressed by appropriate powers
of $1/M$. We thereby arrive at the estimates
\begin{equation}
\label{hard}
\Delta \tilde{g}(M), \Delta g^2_{D}(M) \sim \frac{V_0}{M^4}.
\end{equation}
We can use these as the initial conditions for our RG analysis,  
identifying $m_0 \equiv M$. 

The complete absence of UV SUSY can be identified with the choice $V_0 \sim 
M^4$, but more generally we can consider $V_0 < M^4$, in which case we see 
that the initial conditions of our analysis can easily be highly 
supersymmetric, with further supersymmetric focussing in the IR. 
The SUSY breaking terms 
in the gauge and strongly-coupled matter sectors
are ``hard'' because there are no highly relevant operators allowed, whose 
coefficients would be ``soft terms''. Hard breaking of a symmetry can
 in general naturally be 
small when the symmetry breaking spurion is dimensionful, as our case 
illustrates. 
In a weakly coupled matter sector by contrast, one would naturally have soft scalar 
mass terms, $ \sim V_0/M^2$, and SUSY in this sector would be badly 
broken at $\sqrt{V_0}/M$, which could easily be a very large scale.

If SUSY starts as a fundamental spacetime symmetry, we cannot ignore 
the effects of supergravity.
Anomaly-mediated SUSY breaking (AMSB) \cite{amsb} is a supergravity correction that 
typically introduces gaugino masses $\sim \frac{g^2}{16 \pi^2} 
\sqrt{V_0}/M_{Pl}$, although in 
special circumstances this effect can be much smaller \cite{markus1} 
\cite{split1} \cite{split3}.
Even in the more general case the 
 effect can be subdominant to the effects we consider,  by taking 
$M \ll M_{Pl}$. We will choose parameters in this regime when we give 
illustrative numerical estimates later.  
The gravitino mass is of order $\sqrt{V_0}/M_{Pl}$, which we will again 
use in our later estimates to determine the lightest stable SUSY particle 
(LSP).

\section{Mass Gap in the Strong Sector}

The treatment of accidental SUSY in gauge theory is both incomplete as is and 
not directly applicable to the real world. Our linearized treatment of 
$\Delta g_D^2$ is incompatible with flowing arbitrarily into the IR because 
of the blow-up due to $\gamma_D >0$ in Eq. (\ref{deltasol}). Even if we could 
keep flowing into the IR, SCFT behavior is not compatible with the SM 
physics we eventually want to recover in the IR. Both issues can be 
resolved by arranging for the 
 matter sector to be   
(approximately) superconformal down to a finite scale $\Lambda_{comp}$, 
at which the strong dynamics produces composite states. Let us first see 
the simplest way to do this.

\subsection{Generating $\Lambda_{comp}$}

The generation of a mass gap in the CFT sector is AdS/CFT-dual to ``radius'' 
stabilization in RS1, originally achieved by the Goldberger-Wise mechanism 
\cite{gw}.
 Here we will 
discuss a SUSY version of this mechanism \cite{okada} 
in the dual (S)CFT description, 
and then the effect of high-scale SUSY breaking. The discussion is 
similar but not identical to that of Ref. \cite{markus2} 
The starting point is to assume that the SCFT has an {\it approximate}
 moduli space. (For example, the moduli space could become exact only in the 
large-$N_{CFT}$ limit.) 
We consider minimally a single SUSY chiral superfield modulus, $\omega$.
Conformal invariance is only intact at the origin of moduli space,
while a non-trivial vacuum expectation value (VEV) 
$\langle \omega \rangle$ (spontaneously)
 breaks it. 
We must deviate from the exact 
SCFT in order to stabilize such a VEV $\langle \omega \rangle = 
\Lambda_{comp}$.

Let us introduce a new SCFT operator,  a gauge-singlet chiral primary,
${\cal O}$, with scaling dimension $3 + \epsilon$, where $\epsilon \ll 1$ 
(in practice, $\epsilon \sim 1/10$). We can write a SUSY deformation
of the SCFT in the UV,
\begin{equation}
{\cal L}(M) = {\cal L}_{SCFT} + c \int d^2 \theta ~ 
\frac{\cal O}{M^{\epsilon}} 
+ {\rm h.c.}
\end{equation}
We can match its effects to the $\omega$ effective field theory below 
$\Lambda_{comp}$,
\begin{equation}
{\cal L}_{eff} = \int d^4 \theta |\omega|^2 + \int d^2 \theta ~(\lambda 
\omega^3 
+ c \kappa \frac{\omega^{3+\epsilon}}{M^{\epsilon}}) + {\rm h.c.},
\end{equation}
where $\kappa$ represents a strong interaction matrix element arising in 
matching the deformation. The other terms are superconformally invariant. If 
$\lambda =0$, $\omega$ would be an exact modulus of the SCFT, but we are 
taking it as only approximate, so $\lambda$ is small but non-zero. 
Clearly, the SUSY vacuum satisfies
\begin{equation}
\Lambda_{comp} \equiv \langle \omega \rangle \sim (\lambda/c \kappa)^{1/\epsilon} M,
\end{equation}
which can naturally account for a large hierarchy from $M$ down to 
$\Lambda_{comp}$.

When we consider UV SUSY breaking there is a new subtlety, since now 
we can perturb the SCFT by the lowest component of ${\cal O}$, 
which we will denote by ${\cal O}_|$,
rather than the highest component $\int d^2 \theta ~{\cal O}$. Such a 
highly relevant deformation at maximal strength,
\begin{equation}
{\cal L}(m_0) = {\cal L}_{SCFT} + M^{1- \epsilon} {\cal O}_| + {\rm h.c.},
\end{equation}
would completely destabilize the SCFT and accidental SUSY. 
However, our SUSY-breaking spurion analysis of the last section  
generalizes to this case, and gives 
\begin{eqnarray}
{\cal L}(M) &=& {\cal L}_{SCFT} + V_0/M^{3+ \epsilon} {\cal O}_| + {\rm h.c.}
\end{eqnarray}
The resulting soft scale of SUSY breaking,
 $(V_0/M^{3+ \epsilon})^{1/(1-\epsilon)}$, can naturally be 
small compared to all other effects we study. 

Alternatively, we might be interested in the case of no SUSY in the UV, 
which we explained is equivalent to $V_0 \sim M^4$. In this case, $R$-symmetry 
can be used instead to forbid the SUSY breaking coupling ${\cal O}_|$, 
while permitting the SUSY preserving one, $\int d^2 \theta ~{\cal O}$.
Since ${\cal O}$ has scaling dimension $3+\epsilon$, it has 
superconformal $R$-charge $2 + 2 \epsilon/3$, so even 
$\int d^2 \theta ~{\cal O}$ breaks $R$-symmetry. We therefore assign its 
coefficient $c$  a spurious $R$-charge of $-2 \epsilon/3$. 
Similarly, the ${\cal O}_|$ coupling must have spurious 
$R$-charge $ -2 -2 \epsilon/3$, and therefore can naturally have strength
$\sim c^{3/(\epsilon + 1)}$. Even modestly small $c$ is compatible with
very strongly suppressed SUSY-breaking ${\cal O}_|$ coupling.

In summary, we have a supersymmetric 
mechanism at hand that explains the emergence 
of the IR mass gap in the strong sector, $\Lambda_{comp}$.  
It requires a new supermultiplet of 
operators of the CFT with ${\cal O}(1)$ scaling dimensions. But it is natural 
for significant SUSY breaking not to be communicated via these operators. 
Closely analogous operators (or their AdS/CFT duals)  appear in string 
constructions of warped compactifications compatible 
with high scale SUSY breaking. See the discussion of Ref. \cite{shamit}.

\subsection{Composite Scalars of Accidental SUSY}

 Most of these composites of the strong dynamics will have
masses set by $\Lambda_{comp}$, but a few may be light.
The accidental SUSY means that these light composites come
 in supermultiplets, most simply chiral supermultiplets, $\Phi$.
These are in 
general charged under our original gauge group, since the SCFT was weakly 
gauged.  Then the effective theory below $\Lambda_{comp}$ is given by
\begin{eqnarray}
\label{eff}
{\cal L}_{eff} &=& \int d^4 \theta \Phi^{\dagger} e^{gV} \Phi 
+ \int d^2 \theta ({\cal W}_{\alpha}^2 + W_{eff}(\Phi)) 
+ {\rm h.c.} \nonumber \\
&+& \Delta \tilde{g} \sqrt{2} \phi^{\dagger} \lambda \psi_{\phi} + {\rm h.c.} - 
\frac{\Delta g_D^2}{2} (\sum_{\phi} \phi^{\dagger} T^a \phi)^2.
\end{eqnarray}
Here, the first line contains the 
SUSY approximation, including a superpotential coupling between composites
and a SUSY D-term potential for composite scalars $\phi$ upon integrating 
out the gauge scalar auxiliary fields.
The second line however contains the SUSY breaking effects that have survived 
the RG flow, but they constitute hard breaking of SUSY, not soft terms. 
The first hard breaking is a non-SUSY correction to the gaugino coupling to 
the fermionic ``current'', now composed of
charged composites.  The second term is a SUSY-breaking correction to the 
$D$-term potential for charged scalars, the SCFT D-term operators, $D^a$, 
matching onto the composite field D-terms,  
$\sum_{\phi} \phi^{\dagger} T^a \phi$. 

Let us see what the impact of (approximate) accidental SUSY is on 
the natural scale of light composite scalar masses, $m_{\phi}$. 
Even without SUSY, 
compositeness provides a natural mechanism for having light scalars. 
For example, even pure non-SUSY Yang-Mills theory produces light scalar 
glueballs with masses of order  $\Lambda_{QCD}$, which can naturally be 
small. But this is the analog of $m_{\phi} \sim \Lambda_{comp}$. 
We are really inquiring about the 
 natural hierarchy  that can exist between 
$\Lambda_{comp}$ and $m_{\phi}$. Of course, the best known non-SUSY example is 
 a composite Nambu-Goldstone boson, whose masslessness is protected 
by a spontaneously broken symmetry. But such a scalar has only derivative 
couplings. By contrast the accidental SUSY provides us with 
light scalars which can have non-derivative Yukawa couplings described by 
$W_{eff}$. These couplings do not impact scalar masses by the 
non-renormalization theorem. This is essentially the proposal of 
Ref. \cite{partial}, and we have given it in general terms and have 
corrected it by including symmetry 
protection against terms linear in the $D^a$. 

The couplings of the strong matter sector 
to the external gauge fields radiatively contributes to 
the masses of light scalar composites at one-loop order (just as 
QED contributes to the mass of the charged pion of QCD), 
\begin{equation}
\label{naive}
\Delta m_{\phi}^2 \sim \frac{g^2}{16 \pi^2} \Lambda_{comp}^2, 
\end{equation}
the usual quadratic divergence of fundamental scalars replaced by 
$\Lambda_{comp}^2$ for composite scalars.
If there were no gauginos, such contributions would set a limit on the 
natural hierarchy, $m_{\phi} \sim \frac{g}{4 \pi} \Lambda_{comp}$. 
This is the usual behavior in composite Higgs models \cite{comph} 
\cite{a5}.

But the presence of the gaugino means that there is a possible accidental
SUSY cancellation among the gauge corrections to scalar masses. Crucially, 
this cancellation requires the right SUSY coupling of the gaugino to  
the composites, as well as the right SUSY D-term potential among composite 
scalars. The deviations are measured by the
 hard breaking coefficients $\Delta \tilde{g}^2, \Delta g_D^2$.
As we have seen, without SUSY in the UV (or with badly broken SUSY in the 
UV) the hard breaking is big. But substantial running between 
$m_0$ and $\Lambda_{comp}$ can cure that, so that net gauge 
radiative corrections to scalar masses are given dominantly by
\begin{equation}
\label{improved}
\Delta m_{\phi}^2 \sim \frac{\Delta g^2}{16 \pi^2} \Lambda_{comp}^2 
\sim \frac{\Delta g_0^2}{g_0^2} \frac{g^2}{16 \pi^2} \frac{g^2}{g_0^2} 
(\frac{m_0}{\Lambda_{comp}})^{\gamma_D} \Lambda_{comp}^2.
\end{equation}

Let us first consider that we start with
 order hundred percent UV breaking of SUSY relations among 
$g, \tilde{g}, g_D^2$ at $m_0$, and ask what the best case is 
for accidental SUSY to extend the natural hierarchy between 
$\Lambda_{comp}$ and $m_{\phi}$. For this, we need to assume that 
$\gamma_D$ is small enough 
 ($N_{CFT}$ is large enough) that $(\frac{m_0}{\Lambda_{comp}})^{\gamma_D}$ is 
still roughly of order one, and we need to take the gauge coupling to start 
near strong coupling, $g_0^2 \sim 16 \pi^2$. We then see that the gauge 
radiative corrections cancel to within
\begin{equation}
\Delta m_{\phi}^2 \sim (\frac{g^2}{16 \pi^2})^2 \Lambda_{comp}^2.
\end{equation}
This is a gain of an (IR) gauge loop factor relative to Eq. (\ref{naive}), 
without 
the gaugino. It translates to a theoretical maximum natural hierarchy 
\begin{equation}
\frac{m_{\phi}}{\Lambda_{comp}} \sim  \frac{g^2}{16 \pi^2}.
\end{equation}
This is comparable to the improvement in the natural scale separation 
in going from standard composite Higgs models to Little Higgs models 
\cite{lilh}.

In the absence of UV SUSY, 
accidental SUSY has bought two gains, one being the (maximally) 
extra gauge loop factor from cancellations of gauge radiative corrections, 
and the other that large composite Yukawa interactions do not 
impact scalar masses radiatively. Yet there is a limit to 
the natural hiearchy. In phenomenological terms, if
 one finds the light composite scalars then generic composites are at most 
an inverse-loop-factor away (up to possible fine-tuning).

However, if
 there is some degree of SUSY in the UV couplings $g_0, \tilde{g}_0, 
g_{D 0}^2$, then this can extend the natural 
hiearchy further. 
From the perspective of this paper, this case of $V_0 \ll M^4$ is ``too
good'', in that there can be a very large separation between $\Lambda_{comp}$ 
and $m_{\phi}$, as seen by plugging Eq. (\ref{hard}) into Eq. (\ref{improved}).
Then, accidental SUSY protects $m_{\phi}$ at the lowest
energies, but the strong coupling to which it owes its existence is hidden 
at far higher energies, $\Lambda_{comp}$. Up to fine details (still worthy of 
study), the low-energy 
phenomenology resembles the standard SUSY protection of light 
scalars.
The case of no UV SUSY (equivalent to $V_0 \sim M^4$),
investigated above  is more phenomenologically 
interesting precisely because $\Lambda_{comp}$ and $m_{\phi}$ are naturally 
separated, but not arbitrarily so. Nevertheless we have discussed both cases
 for completeness, and as  prerequisites for the next section.

\section{Split SUSY meets Accidental SUSY}

The most interesting interplay between fundamental SUSY, accidental 
SUSY, and compositeness occurs when there are two types of gauged 
matter sectors, one which is strongly coupled and flowing to a SCFT as 
discussed above, and the other which consists of the more familiar
 weakly coupled matter comprising some charged chiral supermultiplets. 
We follow the discussion of Section 3 and specialize to the 
case $V_0 \ll M^4$. This will result in the highly suppressed 
hard SUSY breaking discussed 
there in the gauge and strongly-coupled matter sectors, as well as soft 
mass terms, $m_0^2 \sim V_0/M^2$, for the weakly coupled matter. 
This means that above $m_0$ the gauge theory is highly supersymmetric, but 
below $m_0$ we must integrate out the massive scalars of the weakly-coupled
chiral multiplets, leaving only their fermionic partners, $\psi_{weak}$. 
This large SUSY splitting of matter multiplets, but without gaugino masses,
is the basic Split SUSY set-up, but here we also have the strongly-coupled 
matter sector.  

The effective theory below $m_0$ in totality therefore does not even have 
supersymmetric particle content, let alone SUSY interactions:
\begin{eqnarray}
{\cal L}(\mu < m_0) 
&=& {\cal L}_{SCFT} + \int d^2 \theta ~{\cal W}_{\alpha}^a {\cal W}^{\alpha}_a  
+  g A_{\mu}^a J^{\mu}_a + \tilde{g} \lambda_a \Psi_J^a 
- \frac{1}{2} g_D D^a D^a, \nonumber \\ 
&+& \bar{\psi}_{weak} i \gamma^{\mu} D_{\mu} \psi_{weak}.
\end{eqnarray}
Nevertheless, the leading matching at the $m_0$-threshold
 has $g$'s satisfying the 
SUSY relations 
\begin{equation}
\label{susyg}
g_0^2 \approx \tilde{g}_0^2 \approx g_{D 0}^2,
\end{equation}
 inherited from $M$, with ${\cal O}(V_0/M^4)$ corrections which we neglect.

Despite the new weakly-coupled matter content,
 the IR theory still only has relevant and marginal 
couplings $g, \tilde{g}, g_D^2$. Let us denote the size of the weak matter 
sector by $N_{weak}$, and we assume that 
\begin{equation}
N_{CFT} \gg N_{weak}.
\end{equation}
Therefore 
to leading order in large-$N_{CFT}$ the RG equations are unchanged.
But again, there is a qualitatively important effect that is missed in this 
limit. The RG equations of the previous section preserved the SUSY relations
of the $g$'s, if they were satisfied in the UV. Here, these relations 
are indeed well satisfied at $m_0$, but the 
corrections to the RG flow due to the $\psi_{weak}$, 
which are subleading in $N_{CFT}$,
no longer preserve the SUSY relations among the $g$'s.
Since we are interested in SUSY cancellations in the radiative 
corrections of composite scalars, we must retain these SUSY-breaking RG 
effects.

Adding these corrections, the RG equations are given by
\begin{eqnarray}
\frac{d g^2}{d \ln \mu} &=& (b_{CFT} + b_{weak}) g^4  \nonumber \\
\frac{d \tilde{g}^2}{d \ln \mu} &=& b_{CFT} \tilde{g}^4  \nonumber \\
\frac{d g_D^2}{d \ln \mu} &=& b_{CFT} g_D^4   - \gamma_D (g_D^2 - g^2),  
\end{eqnarray}
where the only terms subleading in $N_{CFT}$ retained are those that 
push us away from the SUSY relations among the couplings. 
Here, 
\begin{equation}
b_{weak} \sim {\cal O}(N_{weak})/16 \pi^2
\end{equation}
describes the effect on gauge-coupling running due to charged 
$\psi_{weak}$ loops. Without their superpartner scalars however, the 
weak sector has no other interaction than its gauge interactions. In particular
it cannot contribute at one-loop order in the $g$'s to the running of 
$\tilde{g}, g_D^2$. 

We deduce the following linearized RG equations for splitting in the $g$'s,
\begin{eqnarray}
\frac{d \Delta \tilde{g}^2}{d \ln \mu} &=& 2 b_{CFT} g^2 \Delta \tilde{g}^2
- b_{weak} g^4  \nonumber \\
\frac{d \Delta g_D^2}{d \ln \mu} &=& (2  b_{CFT} g^2    
- \gamma_D) \Delta g_D^2 - b_{weak} g^4.
\end{eqnarray}
These have solutions,
\begin{eqnarray}
\frac{\Delta \tilde{g}^2}{g^2}  &=&
 \frac{g^2}{g_0^2} \frac{\Delta \tilde{g}^2_0}{g^2_0} + b_{weak} g^2 
\ln(m_0/\mu) \nonumber \\
\frac{\Delta g_D^2}{g^2}  &=&
 \frac{g^2}{g_0^2} \frac{\Delta g_{D 0}^2}{g^2_0} 
(\frac{m_0}{\mu})^{\gamma_D} + \frac{b_{weak}}{\gamma_D}  g^2 
[(\frac{m_0}{\mu})^{\gamma_D} - 1].
\end{eqnarray}

Specializing to the SUSY initial conditions of Eq. (\ref{susyg}), that is 
$\Delta \tilde{g}_0^2,  \Delta g_{D 0}^2 \approx 0$, 
and running down to the mass gap of the strong sector, $\Lambda_{comp}$,
we arrive at
\begin{eqnarray}
\label{partial}
\frac{\Delta \tilde{g}^2}{g^2}  &=&
 b_{weak} g^2 
\ln(m_0/\Lambda_{comp}) \nonumber \\
\frac{\Delta g_D^2}{g^2}  &=& \frac{b_{weak}}{\gamma_D}  g^2 
[(\frac{m_0}{\Lambda_{comp}})^{\gamma_D} - 1].
\end{eqnarray}
This then translates into non-cancelling 
gauge radiative corrections to composite scalars, 
\begin{equation}
\Delta m_{\phi}^2 \sim \frac{\Delta g^2}{16 \pi^2} \Lambda_{comp}^2.
\end{equation}

Thus, despite gauge interactions connecting
 the strongly coupled matter and weakly
coupled fermions over a large hierarchy, it would appear that 
a high degree of SUSY is accidentally maintained in the light composites, 
permitting a substantial hierarchy between $\Lambda_{comp}$ and the mass of 
light composite scalars. But again, it is interesting that 
this hierarchy cannot naturally be too large. This is our extension of 
the mechanism of Partial SUSY.
We shall make more quantitative illustrative 
estimates in Section 7.


\section{Non-SUSY Warped Models from the CFT View}


Let us rapidly review the core
 structure of realistic weakly-coupled non-SUSY models in warped 5D spacetime 
that exploit the RS1 mechanism to solve the hierarchy problem. 
This will pave the way for incorporating  partial SUSY and understanding 
what it implies for the real world. In particular, we will follow 
Ref. \cite{cust} as our non-SUSY prototype. 
It developed out of the 
suggestion that the basic  RS1 warped hierarchy mechanism could operate 
with most of the SM fields propagating in the 5D bulk spacetime \cite{gw2}, 
that  flavor hierarchies from extra-dimensional 
wavefunction overlaps \cite{martin} could arise attractively in the warped 
setting \cite{neutrino} \cite{pomgher} \cite{huber}, and out of
studies of the implications of the host of precision experimental 
tests \cite{precision}.

We use AdS/CFT duality to map  
to purely 4D models of Higgs compositeness.  In the dual 
description, the extra dimension is replaced by a 
 strongly interacting sector with a large-$N_{CFT}$ expansion, which is 
approximately conformal above a scale $\Lambda_{comp}$.
The KK excitations of the 
higher-dimensional description correspond to 
 ``meson'' and 
``glueball'' composites,  
with weak couplings 
between each other, set by $1/N_{CFT}$. The Kaluza-Klein scale, 
$m_{KK}$, characteristic of the KK spectrum is thereby identified with the 
strong dynamics mass gap, $\Lambda_{comp}$, 
and is taken to be several TeV.  The warp factor effects in 5D
translate into strong RG effects.
The hierarchy problem is solved by the Higgs boson being a light 
composite of the strong dynamics,
 captured in 5D by Higgs localization 
at the IR boundary.
 
Of course, the 5D models are non-renormalizable effective field theories. 
Nevertheless they can quantitatively correlate a number of observables.
But they rely on the existence of an appropriate UV completion. In purely 4D
 terms
they require the existence of CFTs with all but a finite set of (minimal 
CFT-color singlet)
primary 
operators having large scaling dimensions. The finite set of low-dimension 
operators are the key components in coupling to particles external to the CFT.


To start the story, we can consider all the particles of the SM, with the 
exception of the Higgs, as being elementary particles outside of the 
strong CFT sector. They couple (dominantly) linearly to the CFT:
\begin{equation}
{\cal L}(m_0) = {\cal L}_{CFT} + {\cal L}_{SM - H} + A_{\mu} J_{CFT}^{\mu}
 + \bar{\Psi}_{i} \psi_{L i}.
\end{equation}
Here, we are specifying the theory at some fundamental UV scale, which we will 
call $m_0$ for later convenience in Section 7. 
The SM gauge bosons, $A_{\mu}$, couple to global symmetry currents 
of the CFT sector (thereby ``gauging'' them), while the chiral 
SM fermions, $\psi_L$ (in all left-handed notation), couple 
gauge-invariantly to fermionic
 composite operators of the CFT, $\Psi$ \cite{partcomp}. 
At lower scales, $\mu$, we get
\begin{equation}
{\cal L}(\mu) = {\cal L}_{CFT} + {\cal L}_{SM - H} + A_{\mu} J_{CFT}^{\mu}
 + (\frac{\mu}{M})^{\gamma_i} \bar{\Psi}_{i} \psi_{L i},
\end{equation}
where the $\Psi$ have some CFT 
scaling dimensions which we write as
$5/2 + \gamma_i$. (The $J_{CFT}$, being  conserved currents 
have dimension 3.) 
At $\mu = \Lambda_{comp}$ we match onto SM effective field theory.
(We will suppress discussion of the generation of this scale by the dual of 
the non-SUSY Goldberger-Wise mechanism \cite{gw} \cite{nima} \cite{zaff}.)
 The couplings of elementary particles to CFT operators match onto couplings 
to the Higgs composite:
\begin{equation}
{\cal L}_{SM}( \mu < \Lambda_{comp}) =  {\cal L}_{SM - H} + |D_{\mu} H|^2 - V(H) 
+ (\frac{\Lambda_{comp}}{M})^{\gamma_i} (\frac{\Lambda_{comp}}{M})^{\gamma_j} Y_{ij}
\psi_{L i} \psi_{L j}
H.
\end{equation}
Here, the $Y_{ij}$ represent non-hierarchical matrix elements of the 
strong dynamics. As can be seen, 
hierarchies in SM Yukawa couplings nevertheless emerge from the RG factors, 
$(\Lambda_{comp}/M)^{\gamma_i}$. 

The case of $\gamma_i < 0$ is subtle, but necessary at least for 
fitting the large top quark Yukawa coupling. In this case
the corresponding CFT coupling to an elementary SM fermion is relevant in the
IR and can become strong just below $m_0$. This drives the original CFT 
to a new CFT, in which that SM fermion is realized as a 
light ``meson'' \cite{contpom}, in a manner similar to Seiberg duality.
In this way, some of the heavier SM fermions 
can, like the Higgs, be thought of as light composites of a strong CFT, 
and only the remaining lighter 
fermions and SM gauge fields are external to the CFT.

Positive $\gamma_i$ clearly corresponds to irrelevant couplings of the CFT 
to SM fermions in the IR, and to 
small SM Yukawa couplings. But here we must have $\gamma_i < 1$ in order that 
hierarchies among Yukawa coupling
 are considerably smaller than the overall hierarchy $m_0/\Lambda_{comp}$. 
 This explains the need for strong coupling in the 4D description,
not just in the IR at $\Lambda_{comp}$, but all the way up to $m_0$.
At weak coupling, we can trust canonical power-counting, and the only 
composite operator with scaling dimension in the vicinity of $5/2$ is 
the product of a scalar and a fermion.
But in the absence of supersymmetry, the presence of a weakly coupled 
scalar field in the CFT sector would be unnatural. 

Warped models of the type described are in a sense highly successful and
attractive in correlating a great deal of qualitative information, 
in particular the appearance of the electroweak hierarchy and Yukawa or flavor
hierarchies. This is correlated with a suppression of new physics contamination
in the lightest SM particles, which happen to be the best tested. 
In particular it 
alleviates the most stringent constraints of compositeness tests, 
electroweak precision tests, 
flavor-changing neutral currents and 
CP violation. The case of the electroweak $T$-parameter is somewhat 
exceptional, in that satisfying experimental constraints is not automatic
but requires custodial isospin to be taken as an approximate accidental
 symmetry of the CFT 
sector \cite{cust}. 

 But the hierarchy problem is not perfectly solved, precision data still 
require 
the KK mass scale ($\Lambda_{comp}$) to be several TeV\footnote{In particular, 
the array of CP-violation tests provide the most stringent constraints 
\cite{edm} \cite{flavorcp}.}
which acts as the cutoff of the 
effective low-energy SM. The hierarchy between this scale and the weak scale 
continues to pose a non-trivial ``Little Hierarchy Problem''.
This is just where the  improved
 partial SUSY may help, with striking phenomenological 
consequences.

\section{Aspects of Partial SUSY and Realism}

We are deferring until Ref. \cite{future} 
a full and phenomenologically sound 5D warped model, dual to the 
principles discussed in this paper.
 Nevertheless we would like to provide 
enough of a sketch  to suggest the viability and interest of the 
scenario. In particular we want to estimate how much of
 a little hierarchy between the weak scale and $\Lambda_{comp}$ is natural 
with partial SUSY.

\subsection{Basic qualitative features}

In the scenario of Section 5 applied to the real world,
  SM gauge fields are  accompanied by gauginos, 
the weakly-coupled fermions $\psi_{weak}$ 
(with corresponding $\gamma_i > 0$)
are accompanied by sfermions, and the strong CFT is really a 
strong SCFT whose composites, the Higgs and some of the heavy SM fermions, 
also come in complete supermultiplets, $\Phi$. 
The linear couplings of $\psi_{weak}$ to SCFT operators are (slightly) 
irrelevant and therefore, while they are important for the generation of 
hierearchical effective SM Yukawa couplings, they are subdominant from 
the viewpoint of SUSY breaking corrections to the Higgs mass. As long 
as $\Phi = t_R, (t_L, b_L), H_u, H_d$ are all composite chiral 
supermultiplets of the SCFT, the top Yukawa coupling can arise from 
the effective composite superpotential of Eq. (\ref{eff}) with minimal impact 
on Higgs mass. For simplicity here we will take the $\psi_{weak}$ to 
consist of just the first two generations and the $\Phi$ to consist of 
the Higgs multiplets and the entire third generation. (This will leave the 
smallness of $m_b, m_{\tau}$ unexplained, but we will do a more careful job in 
Ref. \cite{future}.) 

We must also outfit our theory in the UV with a protective symmetry 
against the appearance of terms linear in the 
$D^a$ associated to the abelian hypercharge group. 
Yet the MSSM itself does not possess such  a discrete or continuous 
symmetry.   For example, charge 
conjugation invariance is broken in a chiral theory such as the MSSM. 
We will proceed by extending the SM gauge group in the UV to 
 $SU(3) \times SU(2)_L \times SU(2)_R 
\times U(1)_{B-L}$, augmented by the discrete ``left-right'' symmetry under 
which one exchanges the $SU(2)_L$ and $SU(2)_R$ quantum numbers and 
simultaneously charge conjugates with respect to $SU(3) \times U(1)_{B-L}$ 
\cite{lr}. 
The discrete symmetry protects against $D_{B-L}$ terms in the effective 
theory above $\Lambda_{comp}$, and SM generations (including 
SM-sterile right-handed neutrinos) fill out complete representations. 
Below $\Lambda_{comp}$ the extended gauge structure serves no
protective purpose since there is no strong-coupling/warped
``miracle'' such as we have detailed above. We therefore take the
safest route and decouple the extra gauge structure as soon as its job
is done, by 
assuming (naturally enough) that
the extra gauge bosons acquire masses of order $\Lambda_{comp}$.

Until now we have not discussed an origin for gaugino masses protected by 
$R$-symmetry. We do not want them to be massless or we would already 
have seen them. But we do need them to be light enough to 
play a role in safeguarding composite scalar masses. The most natural 
possibility is that the $\Lambda_{comp}$ scale at which conformality is 
broken in the strong sector is accompanied by a moderately lower 
threshold at which accidental SUSY is broken in the strong sector, 
and communicated by $\tilde{g}$ to give gaugino masses. Again, the 
modelling of this within warped compactifications will be deferred to 
Ref. \cite{future}. 

The issue of custodial isospin symmetry in the Higgs sector and our 
experimental measure of violations given by the electroweak $T$-parameter 
is different in partial SUSY compared with completely non-SUSY theories 
\cite{tom}. 
The non-SUSY case is compatible with a single SM-like Higgs multiplet, 
so that in particular the Higgs VEV automatically preserves custodial isospin,
even as it break electroweak symmetry. In the partial SUSY case, the Higgs 
sector is minimally MSSM-like and 
involves two Higgs multiplets. Custodial isospin is generally violated for 
unequal Higgs VEVs. This poses an extra danger for the $T$-parameter. But this 
danger is offset by the higher compositeness scale,, $\Lambda_{comp}$, 
that naturally arises in partial SUSY. The higher scale suppresses new physics 
contributions to $T$. We will study the issue more carefully in Ref. 
\cite{future}. 

A final important phenomenological
consideration, which we will also defer to Ref. \cite{future} 
is precision gauge 
coupling unification, based on
a variation of the result of Ref. \cite{gut} appropriate to 
partial SUSY.

\subsection{Numerical Estimates}

Let us see how natural a Higgs mass of say $200$ GeV is if 
 $\Lambda_{comp} \sim 10$ TeV. Such a high $\Lambda_{comp}$ 
 gives a relatively safe suppression of 
virtual composite effects on the host of precision compositeness, electroweak,
flavor and CP tests. On top of these effects there are also
flavor-changing and CP-violating effects mediated by the
extra elementary gauge bosons of the left-right symmetric model. But 
extra gauge boson masses of order $\Lambda_{comp} \sim 10$ TeV are
roughly adequate for evading the tightest constraints from the kaon
system \cite{rabi}. (A more precise account of these constraints in the
present context will again require more detailed model-building in the
infrared, and is deferred.)

  The leading 
corrections to Higgs masses 
come from the $SU(2)_L$ gauge corrections and from top-stop 
loops. The latter in turn depends on the SUSY breaking stop mass 
due to $SU(3)$ gauge corrections. We therefore specialize to the case 
of external gauge group $SU(n = 2,3)$ and first consider leading radiative 
corrections to composite scalar masses. These are dominated by 
$\Delta g_D^2$ since it grows fastest in the IR due to the slight 
relevance of $D^a D^a$, captured by the exponent $\gamma_D > 0$.
In the IR effective theory $\Delta g_D^2$ gives the SUSY breaking 
correction to the D-term quartic interaction of composite scalars. 
At one-loop order, this hard breaking gives rise to an uncancelled quadratic 
divergence, cut off only by compositeness,
\begin{equation}
\Delta m_{\phi}^2 
= \frac{n^2 -1}{2n} \frac{\Delta g_D^2}{16 \pi^2} \Lambda_{comp}^2.
\end{equation}
Plugging in Eq. (\ref{partial}) for $\Delta g_D^2$ we get
\begin{equation}
\label{bottomline}
\Delta 
m_{\phi}^2 = \frac{n^2 -1}{2n} \frac{g^4}{16 \pi^2} \frac{b_{weak}}{\gamma_D} 
[(\frac{m_0}{\Lambda_{comp}})^{\gamma_D} - 1] \Lambda_{comp}^2.
\end{equation}

Here is an illustrative set of numbers. Let us suppose that 
$m_0 \sim \sqrt{V_0}/M = 10^4$ 
TeV gives the scale of first two generation sfermion masses. 
They are so heavy that no precise flavor-degeneracy is needed to be consistent 
with bounds on their virtual contributions to 
 flavor-changing neutral currents. This situation deserves the name Split 
SUSY, and without strong interaction miracles this splitting would 
give extremely large Higgs mass corrections at two-gauge-loop order.
The last parameter we need to specify is 
$b_{CFT}$ for $SU(n = 2,3)$. We choose $b_{CFT}^{(n=2)} = 1/5$, 
$b_{CFT}^{(n=3)} = 1/10$.  (Recall that the $b_{CFT}$'s are formally of order 
$N_{CFT}/(16 \pi^2)$ in large-$N_{CFT}$ counting.) We take the gauge couplings
at scales of several TeV to be $g_2 \approx 0.6, g_3 \approx 1$. 
Using the leading-in-$N_{CFT}$ running as just a crude estimate we find that 
$g_{2 0} \approx 1, g_{3 0} \approx 2$, far from Landau poles.
 Using Eq. (\ref{gamma})
 we get 
\begin{eqnarray}
\gamma_D^{(2)} \approx 1/12 \nonumber \\
\gamma_D^{(3)} \approx 1/4.
\end{eqnarray}

It is straightforward to check that for two SM generations of 
$\psi_{weak}$, 
\begin{equation}
b_{weak}^{(2)} = b_{weak}^{(3)} = \frac{1}{3 \pi^2}.
\end{equation}
Plugging into Eq. (\ref{bottomline}), 
we find SUSY-breaking stop mass corrections
\begin{equation}
\Delta m_{stop}^2 \approx (700 {\rm GeV})^2.
\end{equation}
This naturally allows stop masses $\sim 700$ GeV, light enough 
not to destabilize a $200$ GeV Higgs via top/stop loops.
Similarly, we find SUSY-breaking electroweak Higgs mass corrections, 
\begin{equation}
\Delta m_{H}^2 \approx (130 {\rm GeV})^2.
\end{equation}
This is also perfectly compatible with a natural Higgs mass of 
 $200$ GeV. Thus all the dominant SUSY breaking 
radiative corrections to Higgs mass are consistent with a naturally light 
Higgs boson, despite having a $10$ TeV compositeness scale. 


Let us check that supergravity effects can be unimportant.
The  anomaly-mediated SUSY breaking contributions to 
gaugino masses can be smaller than $100$ GeV for $M < 10^{14}$ 
GeV, as discussed
in Section 3, taking them out of our consideration. Again from Section 3, 
the gravitino mass is $\sim \sqrt{V_0}/M_{Pl}$, which is $\sim$ TeV with our 
numbers, so that it does not 
constitute the LSP.


\section{Collider Implications}

Even without a fully detailed model, the outlines of the expected 
phenomenology can be given. Well below $\Lambda_{comp}$ we have 
the MSSM but without the sfermion partners of the light fermions. 
They have ``split off'' at the very high scale $m_0$. The LHC phenomenology 
is therefore similar to the scenario of ``More Minimal SUSY'' \cite{moremin}, 
even though the UV physics is quite different. At a hadron collider, 
one will dominantly pair produce the colored 
gluinos and stops among the new particles. 
Assuming effective $R$-parity,
these will decay via real or virtual stops into electroweak superpartners, 
(winos, binos, Higgsinos) plus top or bottom quarks. If the LSP is a
neutralino, such events will be seen as top or bottom pairs plus missing 
energy.  Given sufficiently heavy gravitino as discussed above, 
the LSP may well be a neutralino combination of winos, binos and Higgsinos. 
This can make a suitable WIMP dark matter candidate in standard
fashion. In more detail, it may also be possible to measure the deviations 
of gaugino couplings from gauge couplings, 
$\Delta \tilde{g}$, that reflect the highly 
``split'' nature of the spectrum.


The precise lower bound on the extra gauge boson masses is very
sensitive to CP-violating observables \cite{rabi} and depends in our
case on just how CP violation is implemented. While order 10 TeV
masses appear safe, they are out of range of the LHC. But it is
possible that they are lighter and can be observed as new
$W', Z'$ resonances. Note these are elementary particles not composites of the 
strong sector, in particular
 their couplings to the SM particles are comparable to 
standard electroweak gauge couplings. 

Most spectacularly, there are also composite states of the strong sector, 
AdS/CFT dual to KK excitations of the SUSY SM in the language of
 warped compactification, but their mass scale 
$\Lambda_{comp}$ is expected to be at several TeV. Some of 
these may be visible at the LHC if we are lucky. They would certainly dominate
the physics of the next generation of high energy colliders.

\section*{Acknowledgements}

The author is grateful to Thomas Kramer for early collaboration on 
the subject of partial SUSY, and for helpful 
discussions with
 Kaustubh Agashe, Zackaria Chacko, 
Shamit Kachru, David E. Kaplan, 
Can Kilic, Markus Luty, Rabindra Mohapatra, 
Takemichi Okui, Leanardo Rastelli, Riccardo Rattazzi, 
Matthew Strassler, Sandip Trivedi and Mithat Unsal. 
The author is grateful for the opportunity to air these ideas 
early in their development
 at the ``Shifmania'' workshop in honor of Mikhail Shifman's sixtieth birthday.
The author also thanks the 
University of Maryland Center for Fundamental Physics for its hospitality 
while some parts of this work were being completed. 
This research was
 supported by the National Science Foundation grant NSF-PHY0401513 and 
by the Johns Hopkins Theoretical Interdisciplinary Physics and Astrophysics 
Center.

\end{document}